# Quantum-enhanced phase interrogation of a fiber-optic strain sensor

ROMAIN DALIDET[1,†], MARIALUISA CAPEZZUTO[2,†], HUGO DEGERT[1], DAVIDE D'AMBROSIO[2], SAVERIO AVINO[2], ANTHONY MARTIN[1], SEBASTIEN TANZILLI[1], LAURENT LABONTE[1], PAOLO DE NATALE[3], AND GIANLUCA GAGLIARDI[2,*]

[1]Institut de Physique de Nice, Université Côte d'Azur-CNRS UMR7010, 17 rue Julien Lauprêtre, 06200 Nice, France
[2]Consiglio Nazionale delle Ricerche, Istituto Nazionale di Ottica (INO), via Campi Flegrei, 34 Comprensorio A. Olivetti, I-80078 Pozzuoli, Italy
[3]Consiglio Nazionale delle Ricerche, Istituto Nazionale di Ottica(INO), Largo E. Fermi 6, 50125 Firenze , Italy
[†]These authors contributed equally to this work.
*gianluca.gagliardi@ino.cnr.it

**Fiber-optic sensors enable high-sensitivity measurements of various physical parameters. Here, we demonstrate a quantum-enhanced interrogation of an all-fiber-optic Fabry-Pèrot resonator, employing entangled two-photon interference to probe directly strain-induced optical phase shifts. The resonator is interrogated by a narrow-linewidth laser that is frequency-locked to its resonance mode, thereby exploiting the cavity's inherent sensitivity to variations in the fiber length. To access the phase fluctuations transferred to the interrogation field due to external perturbations on the sensor, we implement an in-fiber Mach–Zehnder interferometer readout. This configuration allows the sensor to operate both with classical coherent light and with entangled photon pairs generated via spontaneous parametric down-conversion. Harnessing two-photon entangled states corresponding to an N = 2 N00N state, we experimentally demonstrate a reduction of the measurement noise below the standard quantum limit. These results provide clear evidence of the quantum advantage in enhancing fiber-optic sensors under practical constraints on optical power.**

Fiber-optic sensors have emerged as powerful platforms for high sensitivity measurements of a wide range of physical quantities, including strain, temperature, pressure, and refractive index. Their passive operation, lightweight nature, immunity to electromagnetic interference, and natural compatibility with distributed and embedded sensing architectures make fiber-based sensors particularly attractive for applications ranging from structural health monitoring to environmental and biomedical sensing [1–4 Furthermore, optical fibers are a well-established technology, today widely used in telecommunications and numerous other domains. A particularly interesting class of fiber sensors is that based on fiber Bragg-grating (FBG) [1,5] structures that can operate either as intrinsic micro-cavities [6] or as high-reflectivity mirrors forming high-quality Fabry–Pérot resonators directly integrated in optical fibers [7,8]. When interrogated with narrow-linewidth lasers, such resonators provide ultra-high sensitivity and high bandwidth for strain, vibration and acceleration measurements [8].

However, limitations due to classical noise prevent any further improvement in the performance currently achievable with fiber sensors. The sensitivity of any optical interferometric sensor is ultimately limited by the standard quantum limit (SQL), i.e. the shot noise, which originates from the Poissonian photon-number fluctuations of coherent light and leads to a phase uncertainty scaling as $\delta\phi = 1/\sqrt{N}$, where N is the average number of photons [9–11]. While this limit can be reduced by increasing the optical power or integration time, many realistic sensing and imaging scenarios impose strict constraints on the photon flux and detection bandwidth, preventing further performance gains through classical means [12–14]. Quantum metrology offers an alternative approach by exploiting non-classical states of light to surpass the SQL at fixed photon number, ultimately reaching the fundamental Heisenberg limit (HL), i.e. $\delta\phi = 1/N$ [9–11]. In this regard, entangled photon pairs or other correlated states represent a paradigmatic resource that can provide a measurable quantum advantage under practical experimental constraints and with small photon numbers [15–18]. Beyond sensitivity enhancement, entanglement enables distributed quantum sensing architectures in which spatially separated sensors share non-classical correlations to perform joint parameter estimation [9,10,14,16]. Such approaches naturally support secure sensing protocols, where the integrity of the measurement data is guaranteed by quantum correlations rather than computational assumptions [19,20]. In this context, delegated quantum sensing schemes, fully relying on entanglement, have been proposed, in which a central node distributes entangled states to remote sensors that locally probe physical parameters while security and trust are enforced at the protocol level [21].

In this work, we implement for the first time a hybrid classical–quantum sensing scheme based on a fiber Fabry-Pérot resonator (FFR) as a strain sensor, capable of detecting static deformation and acceleration applied to the intra-cavity fiber. The FFR sensor is interrogated by a narrow-linewidth laser, which is frequency locked to its resonant mode ensuring efficient transduction of strain into optical phase and frequency changes [6]. An external, unbalanced fiber Mach–Zehnder interferometer (MZI) is devised as a phase readout stage, which can be operated either with classical coherent light or with non-classical light generated via spontaneous parametric down-conversion. (SPDC). Using two-photon interference as a minimal non-classical resource, we demonstrate a reduction of the measurement noise below the SQL, highlighting the relevance of quantum-enhanced readout schemes for fiber-optic sensing in power-constrained regimes.

Our system is based on a diode laser (RIO Planex), operating around a 1560-nm wavelength with a 2-kHz emission linewidth, which is phase modulated through its injection current to create frequency sidebands at 27 MHz. The sensing element is a FFR consisting of two 99-% reflectivity FBG mirrors, separated by 130 mm. The FFR reflected radiation is routed through a fiber circulator to a fast photodetector. Its signal is demodulated at the sideband frequency in a radiofrequency mixer to generate a Pound-Drever-Hall (PDH) error signal with a dispersive-like profile across the cavity resonance [22]. The PDH signal is fed back to the laser current via a 10-kHz bandwidth proportional-integral servo that drives its wavelength to track the FFR resonance. As a result, any deformation and vibration affecting the intra-cavity fiber is converted, within the servo bandwidth, into a corresponding variation of the laser frequency. Controlled strain can be applied to the FFR using a piezoelectric (PZT) stretcher attached to its ends.

While the laser is locked to the FFR, an all-fiber unbalanced MZI, whose arms differ in length by 0.333 ± 0.006 m, serves as external readout stage of the phase fluctuations of the FFR interrogation field (Fig. 1). This interferometer can be illuminated either with classical light, i.e. single photons derived from the interrogation laser, or with quantum-correlated photon pairs generated by a nonlinear generation unit (NLG). The latter is pumped by the same laser and relies on two cascaded periodically-poled $LiNbO_3$ (PPLN) crystals: the first one up-converts the 1560-nm laser light to 780 nm via second-harmonic generation, the second one generates time-energy entangled photon pairs at 1560 nm through spontaneous parametric down-conversion (SPDC). Thus, the NLG photons are phase correlated with the pump laser, whose phase may be affected by the FFR resonance during locking. Strain sensing can be carried out either indirectly, monitoring the electronic signal at the locking-loop output, or directly, measuring the optical phase variations at the MZI output. The fiber-optic MZI is stabilized by an active temperature control and equipped with a piezo-electric stretcher acting on one arm in order to maintain a constant mid-fringe interference condition, i.e., the highest and most-linear response to phase variations. The outputs of the fiber interferometer are monitored using superconducting nanowire single-photon detectors (SNSPDs), which enable both single-photon detection and photon-pair coincidence measurements. The detector signals are processed by a time-to-digital converter (TDC) with a 1-ps time resolution for the registration and analysis of coincidence events.

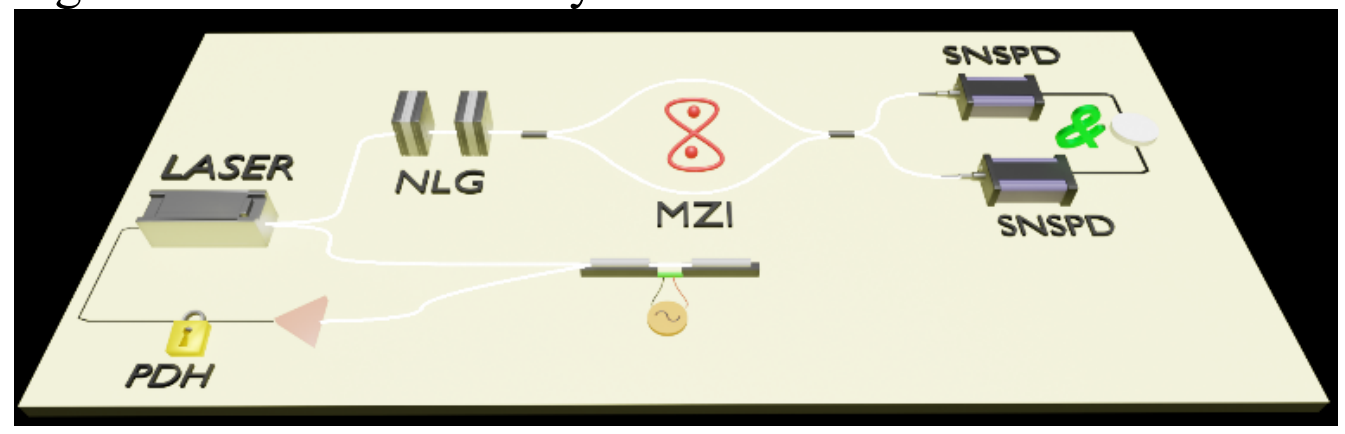

Fig. 1. Experimental setup combining a PDH-locked fiber Fabry–Pérot resonator used as a strain sensor with an unbalanced fiber Mach–Zehnder interferometer (MZI) for phase readout. The interferometer can be operated with either classical light or non-classical light generated by a nonlinear generation unit (NLG): the NLG is formed by two cascaded nonlinear crystals for SPDC pumped with the 1560-nm laser. The MZI transmitted beams are detected with superconducting nanowire single-photon detectors (SNSPDs).

Our main goal is to evaluate how the quantum-light assisted sensor readout compares with its classical counterpart, having the same photon flux at the photodetectors. For this purpose, we send the NLG output light beam, pumped by the laser locked to the FFR, to the MZI and count either single-photon events at the output with a single SNSPD or the photon-pair coincidences of the two output SNSPDs.

In Fig. 2a, we show the time recording of single-photon counting obtained with the TDC over 100 s when a low photon flux beam passes through the MZI. The counting distribution appears to be dominated by Poissonian noise as can be seen from the corresponding Allan deviation plot vs. the integration time, $\tau$ (Fig. 2b). In particular, it is worth noting that the phase fluctuations approach the SQL line with increasing $\tau$ up to ~ 0.3 s, thus confirming that we are mostly limited by the shot noise in the single-photon regime. For longer time scales, the deviation increases likely because of thermal drifts affecting the interrogation laser and MZI.

The fast-Fourier transform (FFT) of the time counting yields the (amplitude) phase noise spectral density (PSD) at the MZI readout with the laser locked to the FFR. When a calibrated strain with root-mean-square (rms) amplitude of 1 nε (1 nε is a fractional fiber length change of $10^{-9}$) is applied to the FFR fiber, sending a 10-Hz sinewave to the PZT, as expected the PSD exhibits a sharp peak emerging from the noise (Fig. 2c). Here, a nearly flat noisefloor is observed for frequencies higher than 1 Hz, in agreement with the white distribution in the time domain. From the measurements shown above, it is worth noting that the PSD level over the flat region from 0.1 to 500 Hz is highly consistent with the shot-noise level calculated from $1/\sqrt{N}$, where $N$ corresponds to the average number of counts at the SNSPD. This points out the intrinsic nature of the noise affecting the system and suggests that other noise sources, e.g., technical and electronic consecutive spectra, when a sinusoidal strain perturbation is applied to the FFR (red dashed line: shot noise).

Fig. 2. Single-photon phase readout with the fiber MZI operated with classical light. (a) Time trace of the single-photon detection counts recorded over 100 s while the interrogation laser is locked to the FFR. (b) Allan deviation of the corresponding phase fluctuations as a function of the integration time (green line: overlapping Allan deviation; red line: shot-noise deviation). (c) PSD retrieved from the 100-s acquisition calculating the FFTs of 10-s single-photon countings, averaged over 10

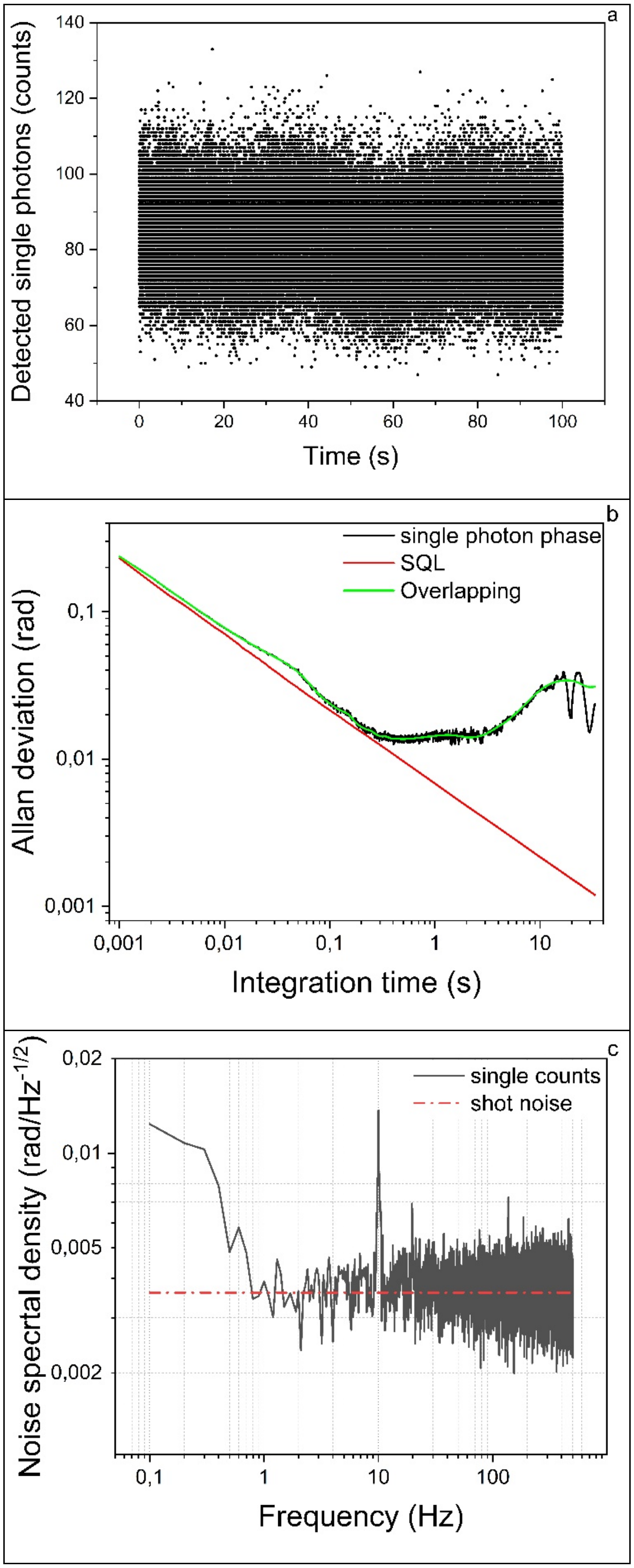

noise, are substantially negligible. In order to further verify the correctness of our analysis, let us consider the Fano factor, defined as [23] $F = Var(N)/N$. This factor quantifies the likelihood to a Poissonian process and, in the case of shot-noise, $F = 1$. Thus, it can be shown for any real PSD that

$$\mathrm{PSD}_{\mathrm{real}} = \sqrt{F}\,\mathrm{PSD}_{\mathrm{shot}} \qquad (1)$$

and similarly, for Allan deviations

$$\sigma_{\mathrm{real}}(\tau) = \sqrt{F}\,\sigma_{\mathrm{shot}}(\tau) \qquad (2)$$

that suggests the deviations exhibit the same slope ($\propto \tau^{-1/2}$) but with a $\sqrt{F}$ offset. In our measurements, the single-photon counting gives $\sqrt{F} \sim 1.1525$, which yields through eq. (1) a PSD ratio of 1.073, in agreement with the average 7.3 % difference observed between the background level of the flat part of the spectrum in Fig. 2c and the shot noise. This is confirmed using the Fano factor in eq. (2) with the corresponding Allan deviations, as can be readily verified from the difference observed in Fig. 2b.

Finally, we carried out strain sensing in a quantum regime, measuring the coincidences of entangled-photon pairs at the MZI readout. In this case, the relevant HL is theoretically rescaled by a factor $1/\sqrt{M}$ with respect to the SQL, $M = 2$ being the entangled-photon number. In Fig. 3a, the coincidences recorded at the MZI output over 100 s are shown: Fig. 4b now points out an Allan stability that is decreased below the shot noise and settles between the theoretical SQL and HL lines calculated for the same photon numbers at varying integration time $\tau$. Indeed, the phase fluctuation is reduced by $\sim \sqrt{1.5}$, for short integration times (i.e., for sampling rate in the audio range), and approaches the shot noise for $\tau \sim 1$ s. This result is confirmed by the PSD spectrum, in Fig. 3c, where a signal-to-noise ratio improvement of $\sim \sqrt{1.5}$ can be appreciated comparing the noisefloor of the spectra recorded for classical and quantum readout around the same strain peaks at 10 Hz.

In conclusion, we experimentally demonstrate a quantum-enhanced all-fiber readout method based on entangled two-photon interference to measure static and dynamic strain sensed by a fiber-optic resonator. By employing a N00N state in a fiber Mach–Zehnder interferometer, we measure the optical phase shift induced by a dynamic strain with a precision below the SQL over a broad range of integration times at low photon-counting rates. A direct comparison with a single-photon interference scheme reveals a sensitivity enhancement that highlights the metrological advantage provided by nonclassical states of light with small photon numbers. The present bound on the achievable strain measurement limit is set by the performance of the cryogenic single-photon detectors, whose efficiency decreases at increasing count rates. Although the current implementation does not yet compete with classical sensing approaches phase stability lies below the SQL (red line) up to τ ~ 1 s, then deviating due to long-term drifts. The blue line represents the calculated HL. (c) PSD retrieved from the 100-s acquisition calculating the FFTs of 10-s coincidences, averaged over 10 consecutive spectra, when a sinusoidal strain perturbation is applied to the FFR (red dashed line: shot noise level, blue dashed line: HL).

Fig. 3. Phase readout with the fiber MZI operated with non-classical light. (a) Time trace of the photon-pair coincidence counts over 100 s while the interrogation laser is locked to the FFR. (b) Allan deviation of the corresponding phase fluctuations as a function of the integration time (green line: overlapping Allan deviation that emphasizes the sub-shot noise performance; red line: shot-noise deviation). The measured

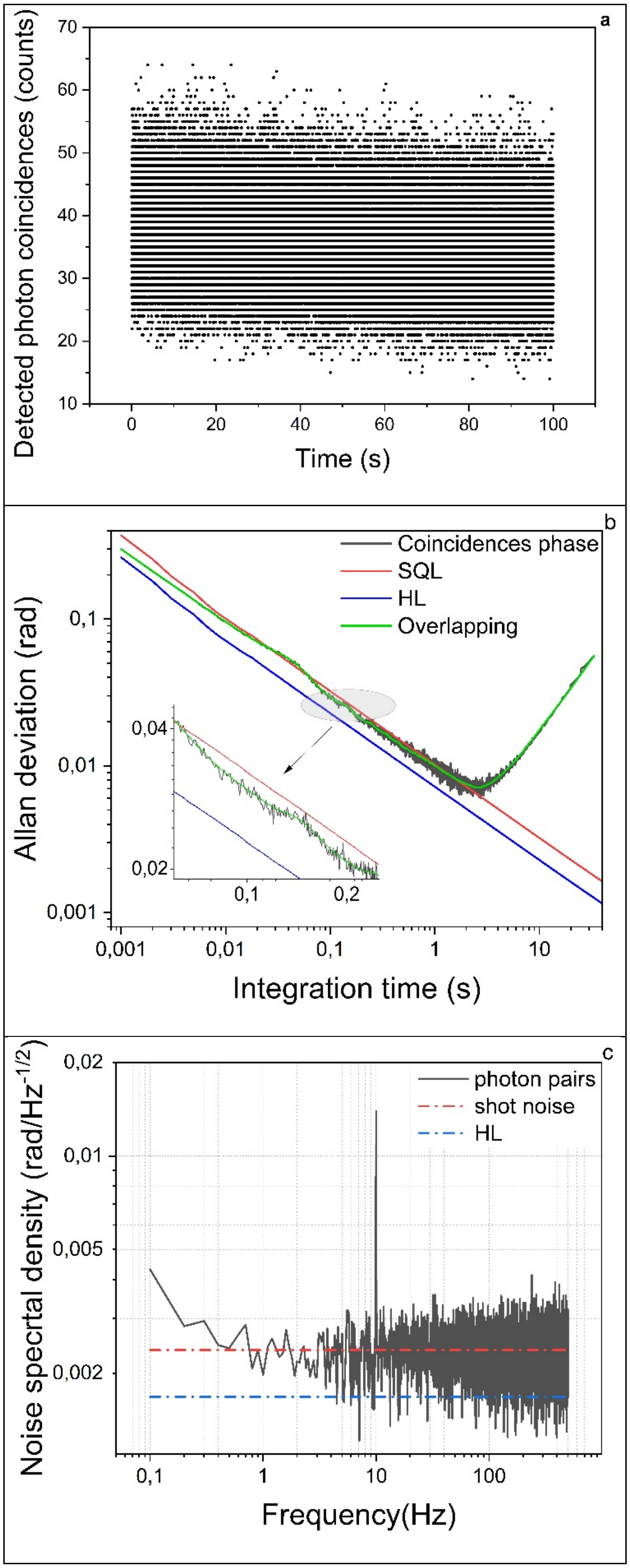

relying on large photon fluxes [7,8], the quantum-enhanced scheme becomes particularly relevant in regimes where the optical power must remain low due to noise constraints, nonlinear effects, or the risk of damaging sensitive samples as well as in other applications where classical high-power interrogation becomes impractical [24-26]. Our results therefore point toward the use of few-photon quantum states to enhance the performance of fiber sensors and more generally of optical resonator sensing platforms.

**Acknowledgments.** This work received funding from the European Defense Fund under grant agreement 101103417 EDF-2021-DIS-RDIS-ADEQUADE. Views and opinions expressed are, however, those of the author(s) only and do not necessarily reflect those of the European Union or the European Commission. Neither the European Union nor the granting authority can be held responsible for them. The authors acknowledge support by CNMS MOST_Spoke12 funded by Italian Ministry of Research under PNRR Program.

**Disclosures**. The authors declare no conflicts of interest.

### Full reference list